\documentclass[a4paper, 12pt]{iopart}
\usepackage{iopams}
\usepackage{color, psfrag, epsfig, graphicx, geometry, wrapfig}
\usepackage{mathrsfs}
\usepackage{url}
\usepackage{multirow}

\usepackage{ulem}
\definecolor{red}{rgb}{1,0,0}

\begin{document}




\title[LIGO-GEO S4 burst search]{First joint
search for gravitational-wave bursts in LIGO and GEO\,600 data}

\author{B~Abbott$^{1}$, R~Abbott$^{1}$, R~Adhikari$^{1}$,
P~Ajith$^{2}$, B~Allen$^{2,3}$, G~Allen$^{4}$, R~Amin$^{5}$,
S~B~Anderson$^{1}$, W~G~Anderson$^{3}$, M~A~Arain$^{6}$
% ...
\\ (LIGO Scientific Collaboration)}
\address{$^{1}$ LIGO - California Institute of Technology, Pasadena, CA  91125, USA}
\address{$^{2}$ Albert-Einstein-Institut, Max-Planck-Institut f\"ur Gravitationsphysik, D-30167 Hannover, Germany}
\address{$^{3}$ University of Wisconsin-Milwaukee, Milwaukee, WI  53201, USA}
\address{$^{4}$ Stanford University, Stanford, CA  94305, USA}
\address{$^{5}$ Louisiana State University, Baton Rouge, LA  70803, USA}
\address{$^{6}$ University of Florida, Gainesville, FL  32611, USA}
% ...

\ead{i.heng@physics.gla.ac.uk}

\begin{abstract} \\
We present the results of the first joint search for
gravitational-wave bursts by the LIGO and
GEO\,600 detectors. 
We search for bursts with characteristic central frequencies 
in the band 768 to 2048 Hz in the data 
acquired between the 22nd of February and the 23rd of March,
2005 (fourth LSC Science Run -- S4).
We discuss the inclusion of the GEO\,600 data in the Waveburst-CorrPower
pipeline that first searches for coincident excess power events
without taking into account differences in the antenna responses
or strain sensitivities
 of the various detectors. We compare the performance of
this pipeline to that of the {\it coherent} Waveburst pipeline 
based on the maximum likelihood statistic. This likelihood statistic 
is derived from a coherent sum of the detector data streams that takes 
into account the antenna patterns and sensitivities of the different 
detectors in the network. 
We find that the coherent Waveburst pipeline is sensitive to signals of
amplitude $30-50\%$ smaller than the Waveburst-CorrPower pipeline.
We perform a search for gravitational-wave bursts using both
pipelines and find no detection candidates in the S4 data set
when all four instruments were operating stably.

\end{abstract}

\section{Introduction}

The worldwide network of
interferometric gravitational wave detectors currently includes the three
detectors of LIGO~\cite{LIGOstatusAmaldi}, as well as the
GEO\,600~\cite{geostatus-amaldi6}, Virgo~\cite{VIRGOstatusAmaldi} and
TAMA300~\cite{TAMAstatusGWDAW04} detectors.
The LIGO and GEO\,600 detectors and affiliated institutions form the LIGO
Scientific Collaboration (LSC). 
The LSC has performed several joint operational runs of its detectors.
During the course of the most recent runs, the detectors have reached
sensitivities that may allow them to detect gravitational waves from distant
astrophysical sources.

Expected sources of gravitational-wave bursts include, for example, 
core-collapse supernovae and the merger phase of inspiralling compact 
object binaries. In general, due to the complex physics involved in such
systems, the waveforms of the gravitational wave signals are not well-modelled. 

There are two broad categories of gravitational-wave bursts searches. 
{\it Triggered searches} use information from an external observation, 
such as a gamma-ray burst, to focus on a short time interval, 
permitting a relatively low threshold to be placed on signal-to-noise 
ratio (SNR) for a fixed false alarm probability.
{\it Untriggered searches} are designed to maximise the detection efficiency
for gravitational-wave bursts for data acquired over the entire run 
(spanning weeks or months depending on the run) for a given false alarm 
probability. In general, untriggered searches are designed to
scan the entire sky for gravitational-wave bursts though searches performed for 
a particular sky location (for example, the Galactic Centre) can also come under this
category.

Previous untriggered burst searches performed by the LSC typically consisted of 
a first stage that
identifies coincident excess power in multiple detectors and a second
stage that tests the consistency of the data with the presence of a
gravitational wave signal ~\cite{s2bursts,s3bursts,s4bursts,s2ligo-tama}.
The Waveburst-CorrPower (WBCP) pipeline is an example of such a two-stage 
analysis.
The first stage, performed by Waveburst, involves a wavelet 
transformation of the data and identification of excess power in time-frequency
volumes that are coincident between multiple detectors~\cite{waveburst}.
A waveform consistency test is then performed by the CorrPower algorithm 
which quantifies how well the detected waveforms match each other by using 
the cross-correlation $r$ statistic~\cite{CorrPower,rstat}.
This approach has been used by the LSC to search for gravitational-wave 
bursts in LIGO data acquired during the second through fourth 
Science Runs. 

One should note that this pipeline requires coincident excess power to be 
observed
in all detectors in the network to trigger the waveform consistency test
performed by CorrPower. Furthermore, CorrPower works on the underlying 
assumption that all detectors in the network have similar responses to
the same gravitational wave signal.
This assumption is valid for the LIGO detectors, which have similar 
antenna patterns.
Their strain sensitivities are also similar, though the
two-kilometre interferometer at Hanford is a factor of two less
sensitive than its four-kilometre counterparts.
On the other hand, GEO\,600 has a different orientation on the Earth 
(see Figure \ref{antenna_pattern} and discussion in section 2), 
so that the received signal in this detector is a different 
linear combination of the $h_+$ and $h_\times$ polarisations from that in the 
LIGO detectors.  Furthermore, the GEO\,600 noise spectrum during the fourth LSC 
Science Run, S4, (22nd of February to 23rd of March, 2005) was quite different 
from those of LIGO (see Figure \ref{detector_sensitivity}), with best GEO\,600
sensitivity around 1 kHz.  As a consequence, the approximation of a common 
signal response breaks down for the LIGO-GEO network. For example, a
low-frequency gravitational-wave burst may appear in LIGO but not be evident 
in GEO\,600.  Alternatively, a high-frequency gravitational-wave burst 
may appear more 
strongly in GEO\,600 than in LIGO if it is incident from a sky direction for 
which the GEO antenna response is significantly larger than those of the LIGO 
detectors.  These effects complicate a coincidence analysis of the sort 
employed by the LSC in previous burst searches. Such analyses demand
coincident excitation in all detectors in the network. As a
result, the sensitivity of the network tends to be limited by the 
least sensitive detector~\cite{s2ligo-tama}.

Coherent burst search algorithms have been developed to fold in data from
a network of detectors with different sensitivities and orientations. 
Methods for coherent burst searches were first described 
in~\cite{guersel-tinto} and~\cite{flanagan_hughes}. 
In~\cite{guersel-tinto}, G$\mathrm{\ddot{u}}$rsel
and Tinto have shown for a network of three detectors that a
gravitational wave signal can be
cancelled out by forming a particular linear combination of data from
detectors in the network, producing what is commonly referred to
as the null stream. 
It is now well-known~\cite{rakhmanov} that the approach of G$\mathrm{\ddot{u}}$rsel and Tinto is a special case of maximum likelihood inference.
In~\cite{flanagan_hughes}, Flanagan and Hughes describe a general 
likelihood method for the detection and reconstruction of the two polarisations
of a gravitational wave signal. A modified likelihood method~\cite{ConstraintL} which uses model-independent constraints imposed on the likelihood functional 
is implemented in the coherent Waveburst (cWB) algorithm~\cite{cwb_doc}.
It uses the maximum likelihood statistic, calculated for each point in the
sky, which represents the total signal-to-noise ratio of the gravitational
wave signal detected in the network.
Coincident instrumental or environmental transient artifacts (glitches) that are unlikely to be consistent between the
detectors will usually leave some residual signature in the null
stream, which can be used as a powerful tool for rejection of
glitches~\cite{WenSchutz,NetConsist}.
Recently it was shown that straightforward application of the maximum 
likelihood method to searches of bursts with unknown waveforms can lead 
to inconsistencies and unphysical results~\cite{ConstraintL,mohanty}. 
All these problems occur due to the rank deficiency of the network 
response matrix and therefore can be cured by a suitable regularisation 
procedure~\cite{rakhmanov}.

In this article, we present the first burst search using data from the three LIGO detectors and GEO\,600, acquired during the fourth Science Run of the LSC. 
We present a search for gravitational-wave bursts between 768 and 2048 Hz using both the Waveburst-CorrPower and coherent Waveburst pipelines.
We begin with a brief description of the detectors in section
\ref{instruments} before describing the two methods used
to analyse the acquired data in section \ref{algorithms}. 
We then detail the additional selection
criteria and vetoes in section \ref{dq}.
We present the results of the search in section \ref{results} and compare the detection efficiencies of the two methods. Finally, we discuss our observations in section \ref{discussion}.

\section{Instruments and data\label{instruments}}

Here, we present a brief description of the main features of the
LIGO and GEO\,600 detectors. A more detailed description of the
LIGO detectors in their S4 configuration can be found in 
\cite{LIGOstatusAmaldi}. The
most recent description of the GEO\,600 detector can be found in
\cite{geostatus-amaldi6} and \cite{geostat_gwdaw10}. 

LIGO consists of three laser interferometric detectors at two 
locations in the United
States of America. There are two detectors at the Hanford site, one 
with four-kilometre arms and another with two-kilometre arms, which we refer
to as H1 and H2, respectively. In Livingston, there is one detector
with four-kilometre arms which we refer to as L1.
Each detector consists of a Michelson interferometer with 
Fabry-Perot cavities in both arms. 
The laser light power builds up in these resonant cavities, enhancing
the sensitivity of the detector.
At the input to the interferometer, there is a power-recycling mirror
which increases the stored laser light power in the interferometer.
This reduces the effect of shot noise, allowing for better sensitivity at
higher frequencies.

The GEO\,600 laser interferometric gravitational wave detector has
been built and operated by a British-German collaboration. It is
located near Hannover in Germany and, along with the three LIGO
detectors, is part of the LSC interferometer network.
GEO\,600 is a Michelson interferometer with six hundred metre arms.  The
optical path is folded once to give a 2400\,m round-trip length. 
To compensate for the shorter arm length, GEO\,600
incorporates not only power-recycling, but also 
signal-recycling (SR), which allows the response of
the interferometer to be shaped, and the frequency of maximum
response to be chosen -- the `SR detuning' frequency. 
During the S4 run, a test power-recycling mirror with $1.35\,\%$ 
transmission was installed, yielding an intra cavity power of only 
500\,W. As a result, the sensitivity of GEO\,600 above 500\,Hz 
was limited nearly entirely by shot noise \cite{hild-amaldi6}. The
SR mirror had about $2\,\%$ transmission and the SR detuning frequency
was set at 1\,kHz.
%
An overview of the signal processing and calibration process in S4 is
given in \cite{opti-combine}.

To calibrate the LIGO and GEO\,600 detectors, continuous sinusoidal 
signals are injected into the actuation signals of some mirrors at 
several frequencies.
The resulting displacement is known and used to determine the 
transfer function of the detector to an incoming gravitational wave,
with an accuracy conservatively estimated at $10\%$~\cite{calLIGO,calGEO}. 
For GEO\,600, the demodulated signal from the main photodetector is
recombined using a maximum likelihood method~\cite{recombGEO}.

\begin{figure}[htb]
   \centering
        \epsfig{file=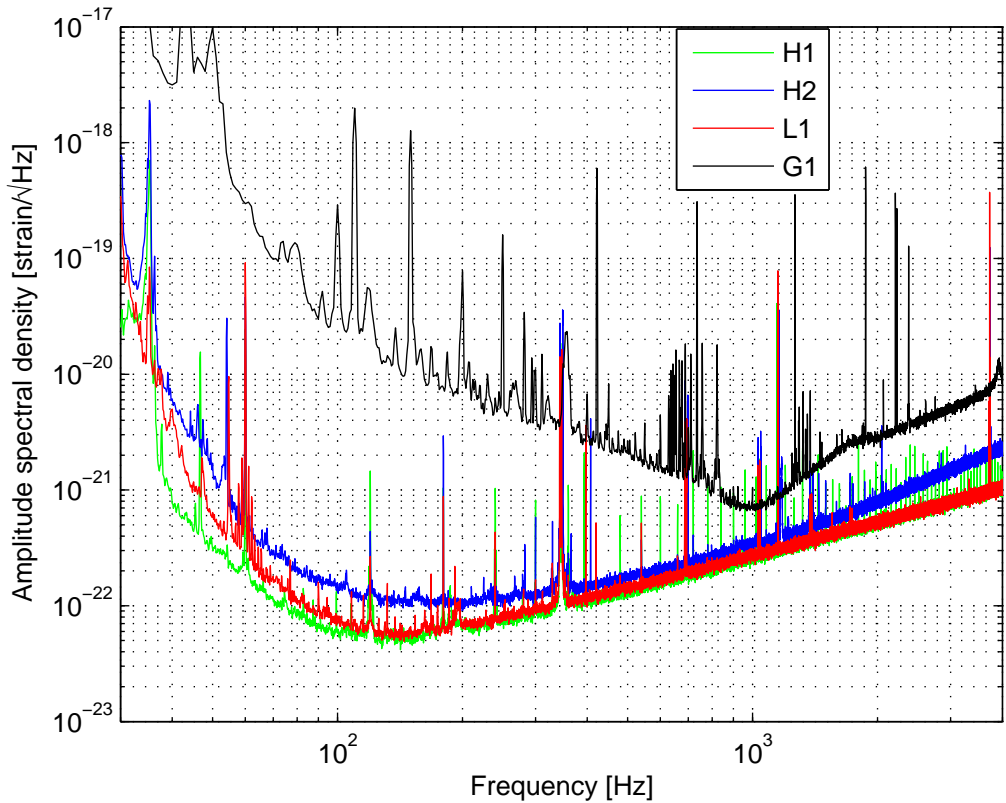, width=\textwidth}
    	\caption{Strain spectral densities of the LIGO Hanford 2-km and 4-km detectors (H1, H2) and
        the LIGO Livingston detector (L1) as well as the GEO\,600 detector (G1)
        during the S4 run. The plotted strain sensitivity curves are the best for the LIGO 
        detectors, obtained on the 26th of February, 2005, for H1 and H2 and the 11th
        of March, 2005, for L1. The GEO\,600 sensitivity curve is typical of the 
        detector's performance during the S4 run.
	}
   \label{detector_sensitivity}
\end{figure}

The strain spectral densities of each detector during S4 are shown in Figure
\ref{detector_sensitivity}. 
The duty factor indicates the percentage of the S4 run each
detector was operational.
GEO\,600 achieved a duty factor of 96.5$\%$,
despite running in a fully automated mode with minimal human
intervention for
operation and maintenance.
H1, H2 and L1 achieved duty factors of $80.5\%$, $81.4\%,$ and $74.5\%$ respectively.

\begin{figure}[htb]
   \centering
        \epsfig{file=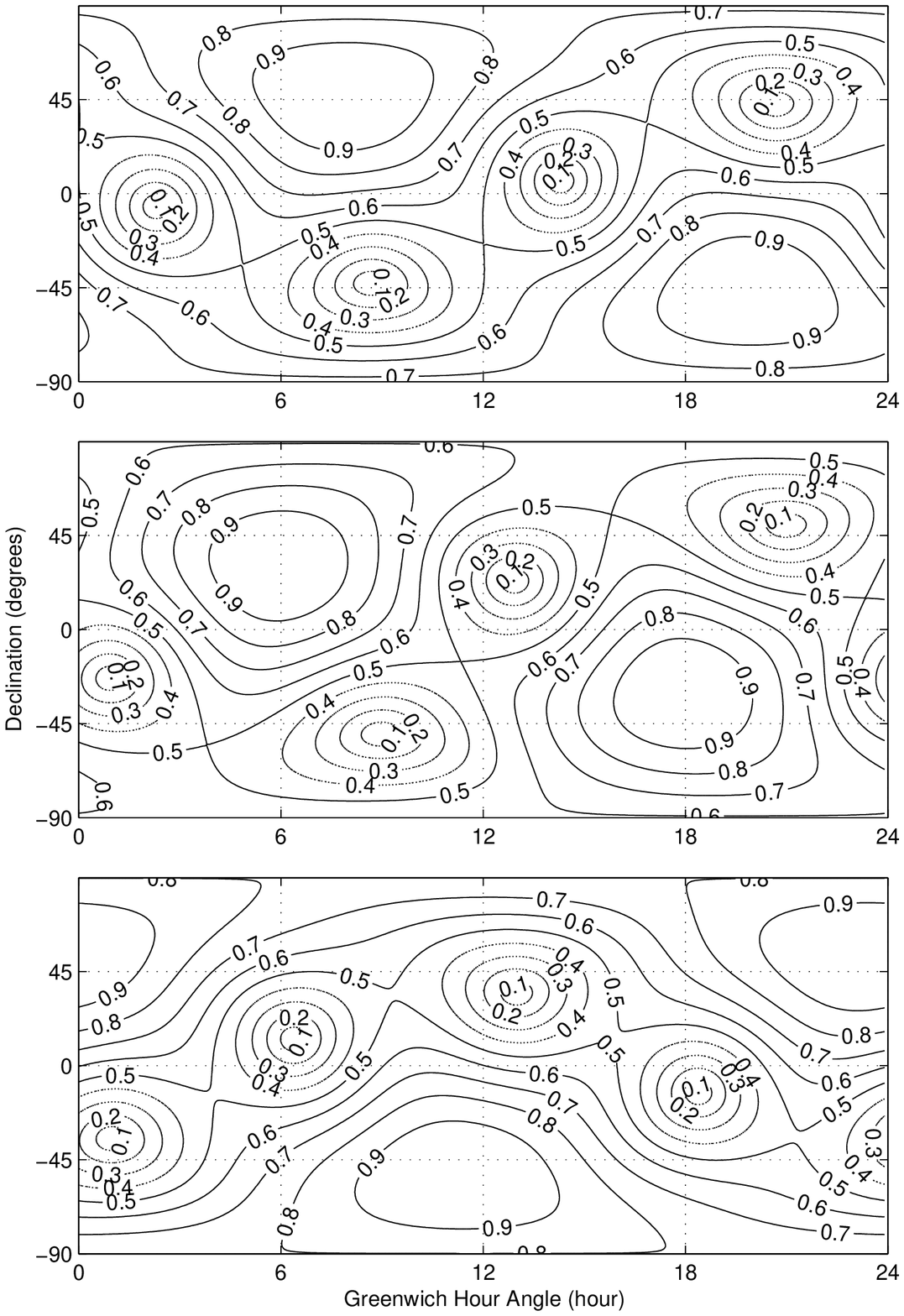, width=0.8\textwidth}
        \caption{Antenna patterns ($F_+^2 + F_\times^2$) of the Hanford (top),
         Livingston (middle) and GEO\,600 (bottom) detectors. The locations of
         the maxima and minima in the antenna patterns for Hanford and
         Livingston are close. However, the antenna pattern for GEO\,600 is
         different from those of the LIGO detectors.
        }
   \label{antenna_pattern}
\end{figure}

The strain detectable at each detector, $h(t)$, for a GW signal
with strain amplitudes of $h_+(t)$ and $h_\times(t)$ in the plus and cross
polarisations, respectively, is given by

\begin{equation}
\label{eqn:hdet}
h(t) = F_+(H_{\rm Greenwich},\delta,\psi) h_+(t) +
F_\times(H_{\rm Greenwich},\delta,\psi) h_\times(t),
\end{equation}

\noindent
where $F_+(H_{\rm Greenwich},\delta,\psi)$ and
$F_\times(H_{\rm Greenwich},\delta,\psi)$ 
are the antenna responses to the plus and cross polarisations. 
The antenna responses depend on the locations and orientations of the 
interferometers on the Earth's surface, where $H_{\rm Greenwich}$ and 
$\delta$ are the Greenwich hour angle and declination of
the source in Earth-centred coordinates and $\psi$ is the
polarisation angle (see \cite{excess_power} for an explicit definition).

Figure \ref{antenna_pattern} shows the sum-squared antenna response
($F_+^2 + F_\times^2$) for each site in the LIGO-GEO network in a fixed-Earth coordinate system. 
The Hanford and Livingston detectors are well aligned to each other and, therefore, have very similar antenna patterns.
On the other hand, the GEO\,600 detector has different antenna patterns,
with peak sensitivities in sky locations that are near the
minima of the LIGO detectors.

\section{Search algorithms\label{algorithms}}

In this section, we describe the two search pipelines used for the analysis.
The WBCP pipeline is almost identical to that used to perform previous 
searches for gravitational-wave bursts~\cite{s2bursts,s3bursts,s4bursts}.
However, for the analysis reported in this article, Waveburst is
applied to data acquired by the LIGO and GEO\,600 detectors, while CorrPower 
is applied only to data acquired by the LIGO detectors 
(see below for further explanation).
The performance of the WBCP pipeline will be compared to 
that of the cWB pipeline.
The same data were processed using the two pipelines.


\subsection{Waveburst and CorrPower pipeline}

We give a brief description of the WBCP pipeline.
More detailed descriptions of the Waveburst and CorrPower algorithms
can be found in~\cite{waveburst} and~\cite{CorrPower} respectively.

The data acquired by each detector in the network are processed
by the Waveburst algorithm which performs a wavelet 
transformation using the Meyer wavelet~\cite{symlet,WBCVS}. This creates
a time-frequency (TF) map of the data. A threshold is applied to this
map to select TF volumes or {\it pixels} with significant excess power.
As with previous LSC GW burst searches, this threshold is set
such that the loudest $10\%$ of the TF pixels are selected.
Coincident excess power pixels from multiple detectors are then
clustered together to form coincident triggers and an overall significance,
$Z_{\rm g}$, is assigned to the coincident pixel cluster~\cite{s4bursts}.

The central time and duration
of these triggers are then passed on to CorrPower~\cite{CPCVS}.
CorrPower calculates the cross-correlation statistic, commonly
denoted by $r$, for the time series data from a pair of
detectors in the following manner:
\begin{equation}
\label{eqn:rstat}
r = \frac {\sum_{i=1}^N (x_i - \bar{x})(y_{i} - \bar{y})}
{\sqrt{\sum_{i=1}^N
    (x_i-\bar{x})^2}\sqrt{\sum_{i=1}^N (y_{i}-\bar{y})^2}} ,
\end{equation}
where $x_i$ and $y_i$ are the $i^{\rm th}$ data sample from the two time series 
from the detector pair, with $\bar x$ and $\bar y$ their
respective means.
The total number of samples over which $r$ is calculated is
denoted by $N$.
This quantity is calculated for a range of time shifts,
corresponding to the range of possible light travel time differences between 
the detectors for gravitational waves incident from different directions 
(up to $\pm 10\,$ms for the LIGO detectors). 
The CorrPower algorithm effectively quantifies
how well the data from different detectors match, thereby
performing an approximate waveform consistency test.

A Kolmogorov-Smirnov test is used to compare the distribution of the
$r$ statistic with a normal distribution with zero mean and a 
variance equal to the inverse of the number of data samples in the time
series. For coincident excess power in multiple detectors, we
expect the $r$ statistic distribution to be inconsistent with a
normal distribution, so we calculate the confidence 
\begin{equation}
\label{eqn:rstat_confidence}
C = -{\rm log}_{10} (S),
\end{equation}
where $S$ is the statistical significance of the $r$ statistic deviation 
from the normal distribution~\cite{rstat}. The overall confidence,
$\Gamma$, is calculated by taking the average confidence for all detector
pairs
\begin{equation}
\label{eqn:rstat_gamma}
\Gamma = \frac{1}{N_{\rm pairs}} \sum_{k=1}^{N_{\rm pairs}} C_k,
\end{equation} 
where $N_{\rm pairs}$ is the total number of detector pairs in the network
(for LIGO, $N_{\rm pairs} = 3$, for LIGO-GEO $N_{\rm pairs} = 6$
but only the 3 LIGO pairs are used here) and $C_k$ is the measured confidence for
the $k^{\rm th}$ detector pair.

The use of CorrPower in this pipeline is best suited to detectors that are
closely aligned, such as the LIGO detectors, since it relies on the
detector responses to incoming gravitational waves to be correlated.
Because GEO\,600 is not aligned with the LIGO detectors, an $r$
statistic calculated for the full LIGO-GEO network would be small
for some sky locations and polarisations for which the detected signal in
GEO\,600 has little or no correlation with the detected signal in LIGO.
This can be accounted for if the source location and signal waveform
are known, but for an all-sky burst search, we find that including
GEO\,600 in the $r$-statistic calculation has little or no benefit.
Therefore, we chose to apply CorrPower to only the LIGO subset of 
detectors.

The search pipeline also performs two diagnostic tests on times when H1 and H2
Waveburst triggers are coincident.
These two tests take advantage of the fact that H1
and H2 are located in the same site and fully aligned.
As a consequence, true gravitational wave signals in H1 and H2 should be 
strongly correlated and have the same strain amplitude.
The pipeline requires, therefore, the H1-H2 triggers to have amplitude ratios
greater than 0.5 and less than 2
(this range is determined by studying the amplitude ratios of simulated
gravitational-wave signals added to the H1 and H2 data streams)~\cite{s4bursts}.
CorrPower also calculates the sign of the cross-correlation between H1 and H2
with no relative time delay, $R_0$, and demands that this quantity be
positive.

\subsection{Coherent Waveburst}

The cWB pipeline uses the regularized likelihood method for the detection 
of gravitational-wave bursts in interferometric data~\cite{ConstraintL}. 
The pipeline is designed to work with arbitrary networks of gravitational 
wave interferometers.
Like the WBCP pipeline described in the previous section,
the cWB pipeline performs analysis in the wavelet domain. 
Both pipelines use the same data conditioning algorithms, but
the generation of burst triggers is different. The WBCP pipeline performs
TF coincidence of the excess power triggers between the detectors.
The cWB pipeline combines the individual detector data streams into a coherent likelihood statistic.

\subsubsection{Regularized likelihood\\\\}

In the presence of a gravitational wave the whitened network output
in the wavelet domain is
\begin{equation}\label{eq:output}
{\bf{w}}={\bf{f_+}}h_+ + {\bf{f_\times}}h_\times + {\bf{n}} \;.
\end{equation}
Here the vectors ${\bf{f_+}}$ and ${\bf{f_\times}}$ characterize the network sensitivity
to the two polarisation components $h_+$ and $h_\times$, and ${\bf{n}}$ is the noise vector.
At each time-frequency pixel [$i$,$j$], the whitened network output is
\begin{equation}\label{eq:vec1}
{\bf{w}} = \left( \frac{a_1[i,j]}{\sigma_1[i,j]},..,\frac{a_K[i,j]}{\sigma_K[i,j]} \right) \; \\
\end{equation}
where $a_1,..,a_K$ are the sampled detector amplitudes in the
wavelet domain, [$i$,$j$] are their time-frequency indices and
$K$ is the number of detectors in the network. Note that the amplitudes $a_k$ take
into account the time delays of a GW signal incoming from a given point in the sky.
In the cWB analysis, we assume that the detector noise is Gaussian and quasi-stationary.
The noise is characterised by its standard deviation $\sigma_k[i,j]$ and
may vary over the time-frequency plane. The antenna pattern
vectors ${\bf{f_+}}$ and ${\bf{f_\times}}$ are defined as
follows:
\begin{equation}
\label{eq:vec2}
{\bf{f_{+(\times)}}} = 
\left( \frac{F_{1+(\times)}}{\sigma_1[i,j]},..,\frac{F_{K+(\times)}}{\sigma_K[i,j]}
\right).
\end{equation}
We calculate the antenna pattern vectors in the dominat polarisation
frame~\cite{ConstraintL}, where we call them $f_1$ and $f_2$.
In this frame, they are orthogonal to each other:
$(\bf{f_1}\cdot\bf{f_2})=0$.
The maximum log-likelihood ratio statistic is calculated as
\begin{equation}
\label{eq:lMax}
L = \sum_{i,j\in{\Omega_{TF}}}{\bf{w}} P {\bf{w}^T}, 
~~P_{nm}=e_{1n}e_{1m}+e_{2n}e_{2m}
\end{equation}
where the time-frequency indices $i$ and $j$ run over some area $\Omega_{TF}$ 
on the TF plane selected for the analysis (network trigger) 
and the matrix $P$ is a projection constructed from the unit vectors ${\bf{e_1}}$
and ${\bf{e_2}}$ along the directions of ${\bf{f_1}}$ and ${\bf{f_2}}$ respectively.
The null space of the projection $P$ defines the reconstructed
detector noise which is often called the null stream. 
The null energy $N$ is calculated by
\begin{equation}
\label{eq:null}
N = E - L ,
\end{equation}
where
\begin{equation}
E = \sum_{i,j\in{\Omega_{TF}}}{|{\bf{w}}|^2}, 
\end{equation}
and $|{\bf{w}}|$ is the vector norm of ${\bf{w}}$.
The null energies $N_k$ for individual detectors can be also
reconstructed~\cite{ConstraintL}.
We also introduce a correlated energy $E_{\rm c}$ which is
defined as the sum of the likelihood terms corresponding to the 
off-diagonal elements of the matrix $P$.

However, the projection $P$ may not always be constructed. For example,
for a network of aligned detectors $|\bf{f_2}|=0$ and the unity vector 
${\bf{e_2}}$ is not defined. As shown in~\cite{ConstraintL}
even for mis-aligned detectors  the network may be much less sensitive 
to the secondary GW component ($|{\bf{f_2}}|<<|{\bf{f_1}}|$) and
it may not be reconstructed from the noisy data.
In order to solve this problem, we introduce a regulator by changing the
norm of the ${\bf{f_2}}$ vector
\begin{equation}
\label{eq:reg}
|{\bf{f'_2}}|^2 = |{\bf{f_2}}|^2
+ \delta \left(|{\bf{f_1}}|^2 - |{\bf{f_2}}|^2\right),
\end{equation}
where the parameter $\delta$ is selected to be 0.1. The regularized
likelihood is then calculated by using the operator $P$ constructed
from the vectors ${\bf{e_1}}$ and ${\bf{e'_2}}$, where
${\bf{e'_2}}$ is ${\bf{f_2}}$ normalised by $|{\bf{f'_2}}|$.
All other coherent statistics, such as the null and correlated energies, are
calculated accordingly.

\subsubsection{Reconstruction of network triggers\\\\}

Coherent Waveburst first resamples the calibrated data streams to
4096 Hz before whitening them in the wavelet domain.
The Meyer wavelet is used to produce time-frequency maps
with the frequency resolutions of 8, 16, 32, 64, 128 and 256 Hz.
An upper bound on the total energy $|{\bf{w}}|^2$ is then 
calculated for each network pixel; if greater than
a threshold, the total energy is then computed for each of
64800 points in the sky placed in a grid with
$1^\circ{\times}1^\circ$ resolution.  If the maximum value
of $|{\bf{w}}|^2$ is greater than 12-13 (depending on the frequency
resolution), the network pixel is selected for likelihood
analysis. The selected pixels are clustered together to form 
network triggers~\cite{cwb_doc}.

After the network triggers are identified, we reconstruct their parameters, 
including the two GW polarizations, 
the individual detector responses and the regularised likelihood
triggers. All the trigger parameters are calculated 
for a point in the sky which is selected by using a criteria
based on the correlated energy and null energy. Namely, we select such a point
in the sky where the network correlation coefficient $cc$ is
maximised:
\begin{equation}
cc = \frac{E_{\rm c}}{N+E_{\rm c}},
\end{equation}
For a GW signal at the true source location a small null energy and large
correlated energy is expected with the value of $cc$ close to unity. 

The identification of the network triggers and reconstruction of their parameters
is performed independently for each frequency resolution. As a result
multiple triggers at the same time-frequency area may be produced.
The trigger with the largest value of the likelihood in the group is selected for
the post-production analysis.

\subsubsection{Post-production analysis\\\\}

During the cWB post-production analysis, we apply additional
selection cuts in order to reject instrumental and environmental
artifacts. For this we use coherent statistics calculated during the production stage.
Empirically, we found the following set of the trigger selection cuts
that perform well on the S4 LIGO-GEO data.

Similar to the regularized likelihood statistic, one can 
define the sub-network likelihood ratios $L_{k}$ where the energy of the reconstructed 
detector responses is subtracted from $L$:
\begin{equation}
\label{eqn:lmax}
L_{k} = L - (E_k - N_k),
\end{equation}
where
\begin{equation}
E_k=\sum_{i,j\in{\Omega_{TF}}}w^2_k[i,j],
\end{equation}
and $w_k[i,j]$ are the components of the whitened data vector 
defined by Eq.~\ref{eq:vec1}.
In the post-production analysis we require that all $L_k$ are greater than
36 which effectively removes single-detector glitches.

Another very efficient selection cut is based on the network correlation coefficient $cc$
and the rank SNR $\rho_k$.
Typically, for glitches, little correlated energy is detected by the network
and the reconstructed detector responses are inconsistent with the
detector outputs, which result in a large null energy: $E_{\rm c}<N$ and
$cc\ll1$.  For a gravitational wave signal, we expect $E_{\rm c}>N$ 
and the value of $cc$ to be close to unity. We define the effective rank SNR 
as
\begin{equation}
\label{eqn:rhoeff}
\rho_\mathrm{eff} = \left( \frac{1}{K}\sum_{k=1}^K \rho_k^2
\right)^{cc/2} ,
\end{equation}
where $\rho_k$ is the non-parametric signal-to-noise ratio for
each detector based on the pixel rank statistic~\cite{rankSNR}
\begin{equation}
y_k[i,j] = -\ln\left( \frac{R_k[i,j]}{M} \right).
\end{equation}
In the equation above $R_k[i,j]$ is the pixel rank (with $R=1$ for the
loudest pixel) and $M$ is the number of pixels used in the ranking
process.
The statistic $y_k[i,j]$ follows an exponential distribution, independent of the
underlying distribution of the pixel amplitudes, $w_k[i,j]$.
The $y_k[i,j]$ can be mapped into rank amplitudes $x_k[i,j]$
which have Gaussian distribution with unity variance.
The $\rho_k$ is calculated as the square root of the sum of $x_k^2[i,j]$ 
over the pixels
in the cluster and it is a robust measure of the SNR
of detected events in the case of non-Gaussian detector noise.
We place a threshold on $\rho_\mathrm{eff}$ to achieve the false
alarm rate desired for the analysis.

\section{Data Quality\label{dq}}

Spurious excitations caused by environmental and instrumental noise
increase the number of background triggers in gravitational-wave burst searches.
Periods when there are detector hardware problems or when the ambient 
environmental noise level is elevated are flagged and excluded from the 
analysis.
These {\it data quality flags} are derived from studies of diagnostic 
channels and from entries made in the electronic logbook by 
interferometer operators and scientists on duty that indicate periods of 
anomalous behaviour in the detector.
Additionally, we veto times when data triggers are observed in
coincidence with short-duration instrumental or environmental
transients. 

To maximise our chances of detecting a gravitational-wave burst, we must 
balance the reduction of each detector's observation time due to data 
quality flags and vetoes against the effectiveness for removing background 
triggers from the analysis.
The data quality flags and vetoes for the LIGO and GEO detectors
are outlined below. 
Out of the 334 hours of quadruple coincidence observation time, 257
hours remained after excluding periods flagged by the data
quality flags.
This observation time is common to both pipelines. The total
livetime of data analysed by cWB is larger by $1\%$ because of
different processing of data segments.

%

\subsection{GEO\,600}

\subsubsection{Data quality flags\\\\}

GEO\,600 data quality flags include periods when the data acquisition
system is saturated (overflow) and when the $\chi^2$ value is too high, 
as explained below.

The GEO\,600 data stream is calibrated into a time series representing
the equivalent gravitational-wave strain at each sample.
The GEO\,600 calibration process determines if the noise, as measured
by the acquired data, is close to that expected from the optical
transfer function by using the $\chi^2$ statistic~\cite{hild-amaldi6}.
If the $\chi^2$ values are too high, it means that the calibration
is not valid.
Therefore, the $\chi^2$ values from the calibration process are an
indicator of data quality.

\subsubsection{Excess glitches\\\\}

During the first 10 days of the S4 run, one of the suspended GEO
components came into contact with a nearby support structure.
This caused GEO data to be glitching excessively between the
22nd of February and the 4th of March, 2005.
The glitch rate fluctuated dramatically over this period because the
distance between the component and the support structure 
changed as a function of temperature.
Given the large variability in the glitch rate (about one order
of magnitude on a timescale of hours), we decided to exclude this period
from the analysis.

\subsection{LIGO}

The data quality flags and auxiliary-channel vetoes used with
the LIGO detectors are explained in more detail in \cite{s4bursts}.
Basic data quality cuts are first applied to LIGO data segments so as to
exclude periods when the detector is out of lock or when simulated GW signals 
are injected into the detector. Additionally, 
data segments are excluded from the analysis when there clearly
are problems with the LIGO hardware or when environmental noise
sources cause spurious transient noise in the data.
%

We rejected periods when injected sinusoidal signals used for
calibration were not present due to problems in the injection
hardware. Since the calibration was unknown for these periods, 
totaling 1203 seconds, the data were excluded from the analysis.
A study based on single-detector triggers showed correlations between
the loudest triggers and the speeds of local winds. This was most
prominent in H2. Therefore, data were not included in the analysis
when the wind speed at the Hanford site was greater than 56\,km/hour 
(35 miles per hour). This excluded a total of 10303 seconds of
four-detector livetime.
Seismic activity between 0.4 and 2.4\,Hz was observed to
cause transients in the detector noise.
Excess coincidences were observed between H1 and H2
when there was elevated seismic activity in this frequency range. 
As a result, time intervals when the root-mean-squared
seismic signal exceeded seven times its median value were excluded
from the analysis. This accounted for 11704 seconds of the
four-detector livetime.
Correlations were also observed between single-detector triggers
and times when data overflows occurred in an
analog-digital-converter (ADC) in the length sensing and control
subsystem. A data quality flag for the data overflows excluded
10169 seconds of four-detector livetime.
Transient dips in the stored light in the arm cavities
were found to be strongly correlated with periods of high single-detector
rates. Data were excluded from the analysis when the change in 
measured light relative to the last second was greater than $5\%$
for H2 and L1. A threshold of $4\%$ relative change was used for H1.

In addition to the exclusion of data segments, triggers attributed to
short-duration instrumental or environmental artifacts are
excluded from the analysis.
This is done by applying vetoes based on triggers generated from 
auxiliary channels found to be in coincidence with
transients in the gravitational wave data, where veto effectiveness
(efficiency versus deadtime) is evaluated on time-shifted background 
data samples prior to use.

\section{Results\label{results}}

Here we present and compare the results of the WBCP and cWB
pipelines applied to the LIGO and GEO\,600 data. 

A total of 257 hours of quadruple coincidence data were processed 
with both the WBCP and cWB pipelines to produce lists of coincident 
triggers, each characterised by a central time, duration, 
central frequency and bandwidth.
In addition to these characteristics, each trigger also has an
estimated significance with respect to the background noise.
Waveburst calculates the overall significance, $Z_{\rm g}$, while CorrPower
calculates the confidence, $\Gamma$.
For coherent Waveburst, each trigger is characterised by the
likelihood and effective SNR (see Eq.~\ref{eqn:lmax} and~\ref{eqn:rhoeff} respectively).
Although WBCP calculates $\Gamma$ using only the LIGO
detectors, for convenience, we will refer to coincident triggers from either
pipeline as {\it quadruple coincidence triggers}. 
The name is still valid for WBCP triggers since the Waveburst stage of the
pipeline requires coincident excess power in all four detectors
in the network.

The central frequencies for triggers from both pipelines were restricted to lie
 between 768 and 2048 Hz. 
This is because the sensitivity of the GEO\,600 detector is closest to 
the LIGO detectors in this frequency range 
(see Figure \ref{detector_sensitivity}). Moreover, the noise of 
GEO\,600 is not very stationary at frequencies below 500 Hz, and 
many spurious glitches can be observed in the acquired data.
CorrPower computes the $r$ statistic over a broader band (64--3152 Hz),
using only LIGO data.

For both pipelines, the L1 data are shifted with respect to H1, H2 and G1 
data by 100 3.125-second time steps.
The applied time shift is sufficiently large that any short gravitational-wave 
bursts present in the data cannot be observed in coincidence in all detectors.
Therefore, we can study the statistics of the noise and tune the thresholds 
of the pipeline without bias from any gravitational wave signals that 
might be present in the data.
The goal of the tuning is to reduce the number of time-shifted
coincidences (background triggers) while maintaining high 
detection efficiency for simulated gravitational wave signals.

The efficiency of the pipeline at detecting gravitational-wave bursts for the
selected thresholds is determined by adding into the data simulated 
gravitational wave signals of various morphologies and amplitudes. For this
study, we used sine-Gaussians, sine waves with a Gaussian envelope, given
in the Earth-fixed frame by
%
\numparts\label{eq:sg}
\begin{eqnarray}
h_+ (t) = h_0 \sin (2\pi f_0 [t-t_0]) \exp[-(2 \pi f_0 [t-t_0])^2 /2Q^2], \\
h_\times (t) = 0,
\end{eqnarray}
\endnumparts
%
%
where $t_0$ and $h_0$ are the peak time and amplitude of the envelope, $Q$ is the
width of the envelope, and $f_0$ is the central frequency of the signal.
The antenna responses (see Eq.~\ref{eqn:hdet}) are generated for
each simulated signal assuming a uniform distribution in the sky
and a polarisation angle $\psi$ uniformly distributed on [0,$\pi$].
The signal strength is parameterised in terms of the root-sum-squared amplitude of the signal, $h_\mathrm{rss}$,
\begin{equation}\label{eq:hdetrss}
h_\mathrm{rss} \equiv \sqrt{ \int  \left (|h_+(t)|^2 +
|h_{\times}(t)|^2 \right ) \, \mathrm{d}t }  .
\end{equation}
The detection efficiency is the fraction of injected signals that produce
triggers surviving the selected thresholds for the respective pipeline.
We characterise the sensitivity of each pipeline by its $h^{50\%}_\mathrm{rss}$, 
which is the $h_\mathrm{rss}$ at which $50\%$ of the injected signals are observed 
at the end of the pipeline (detection efficiency).

\subsection{Waveburst-CorrPower analysis}

\begin{figure}[!htb]
        \centering
        \epsfig{file=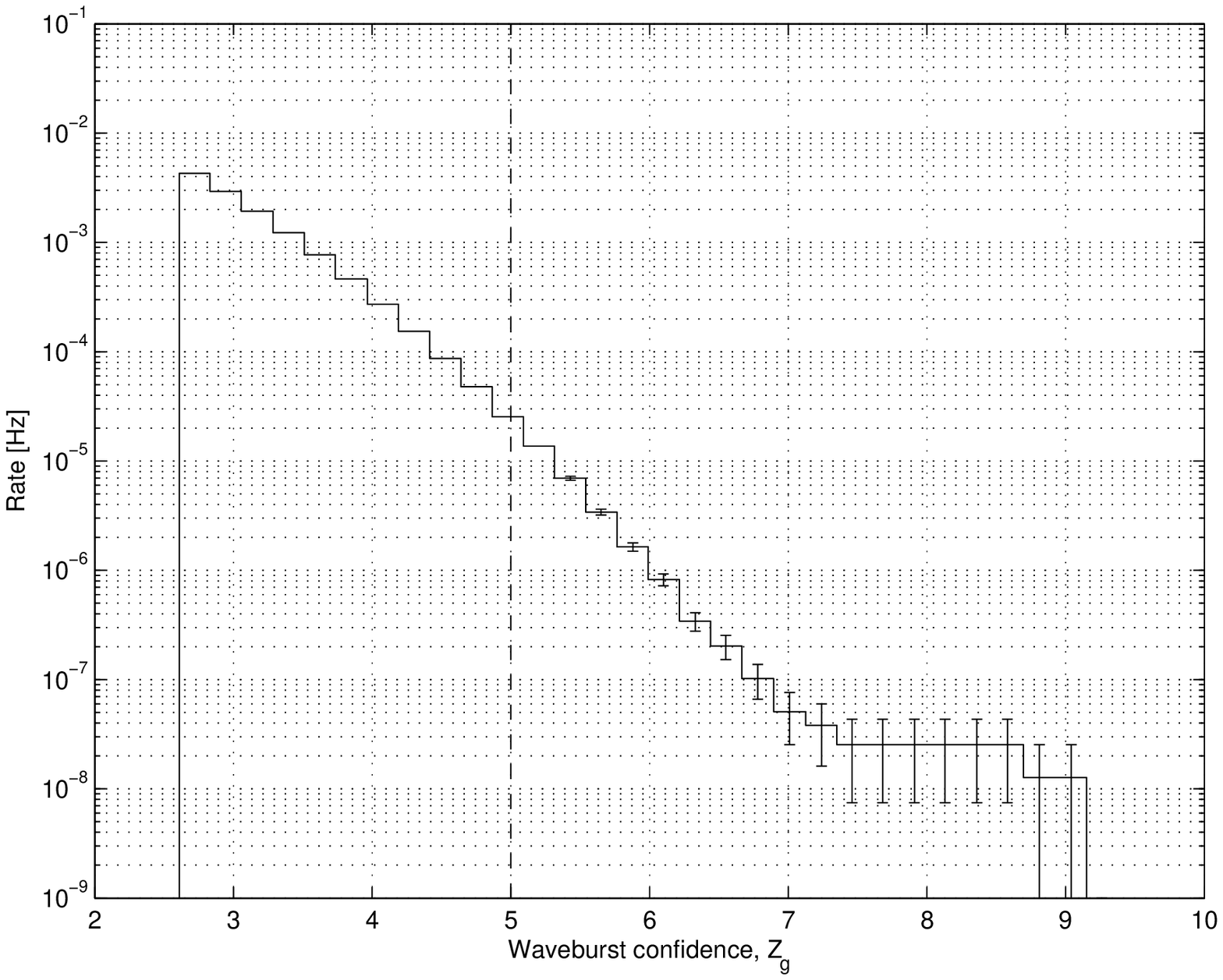, width=\textwidth}
        \caption{Quadruple coincidence rate as a function of the 
         threshold on the Waveburst significance, $Z_{\rm g}$. The threshold
         used for this analysis is indicated by the dashed line. The error bars indicate the
         range corresponding to $\pm \sqrt{n}/T$, where $n$ is
         the number of triggers observed above the $Z_{\rm g}$ threshold over the livetime $T$.
	}
        \label{wb_rate}
\end{figure}

For the WBCP pipeline, there are two threshold values to select.
The background quadruple coincidence rate as a function of the threshold on 
Waveburst significance is shown in Figure \ref{wb_rate}. 
Since the calculation of
the $r$-statistic by CorrPower is computationally expensive and time consuming, we
reduce the number of triggers by selecting a Waveburst significance threshold of $Z_{\rm g}=5$, for a false alarm rate of approximately $3 \times 10^{-5}$ Hz.

\begin{figure}[!htb]
        \centering
        \epsfig{file=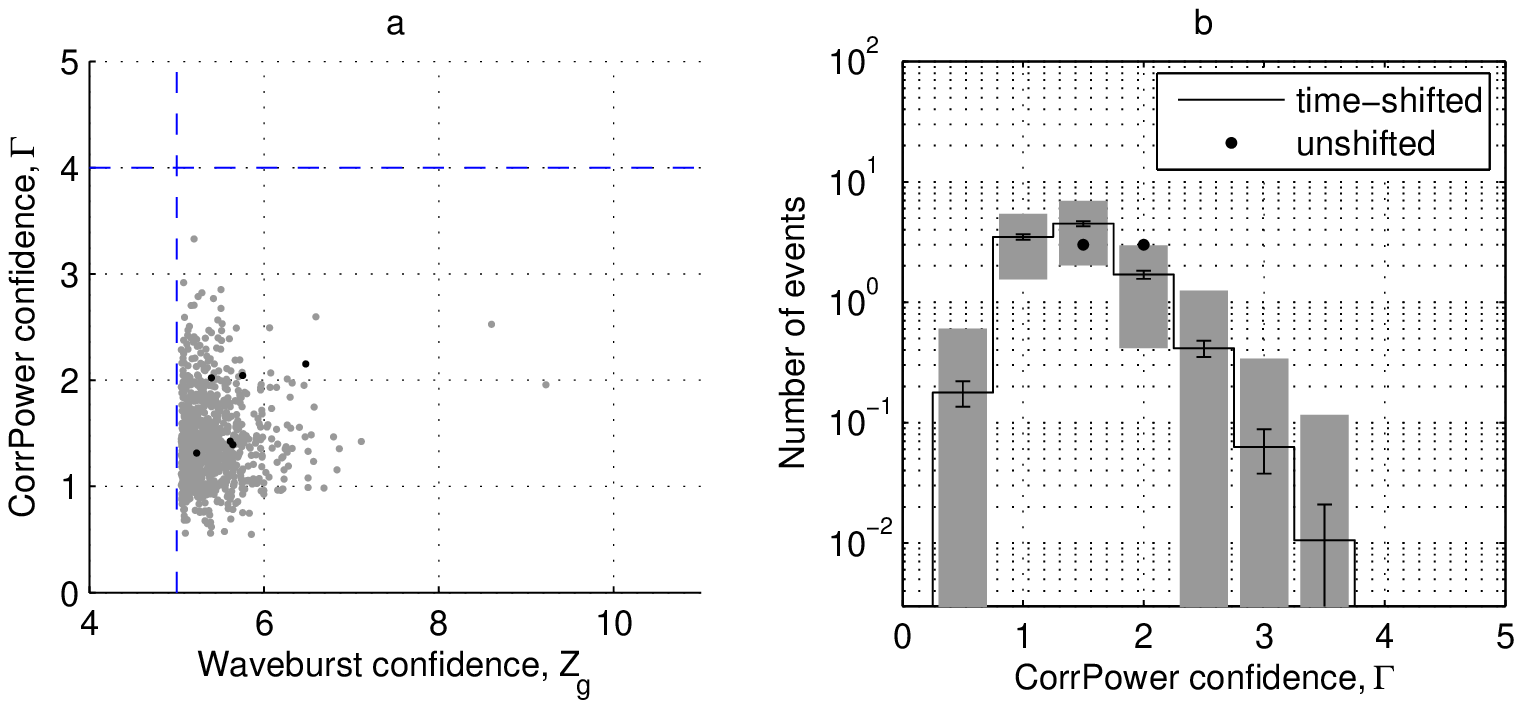,
	 width=\textwidth}
        \caption{(a) Scatterplot of $r$-statistic confidence, $\Gamma$, versus 
        Waveburst $Z_{\rm g}$. 
	The time-shifted background triggers are plotted as grey
	dots while the unshifted triggers are plotted as black
	dots.
	The dashed 
        line indicates the $\Gamma$ threshold chosen for this analysis.
        (b) Overlaid histograms of the unshifted triggers and the
        $\Gamma$ distribution for the time-shifted triggers averaged over the 100 time shifts.
        The grey patches indicate the standard deviation in the number of 
	triggers at each time shift. The error bars indicate the
	range corresponding to $\pm \sqrt{n}/100$, where $n$ is
	the total number of triggers in each bin.
        }
        \label{rstat-scatter}
\end{figure}

The CorrPower confidence, $\Gamma$, is then calculated for each
surviving trigger. 
A scatter plot of $\Gamma$ versus $Z_{\rm g}$ for these triggers
can be seen in Figure \ref{rstat-scatter}a. 
Note that the all triggers have $\Gamma$ values less than 4.
The distributions of the $\Gamma$ values of both the
time-shifted background triggers and unshifted triggers 
are plotted in Figure \ref{rstat-scatter}b. 

Table 1 shows the number of background coincidences and the $h_\mathrm{rss}^{50\%}$ values for sine-Gaussian injections of different central frequencies for several trial values of the threshold on $\Gamma$: $\Gamma > 0$ (CorrPower not used), $\Gamma > 3$ and $\Gamma > 4$.
We note that the $h^{50\%}_\mathrm{rss}$ values for a threshold
of $\Gamma = 4$ are only a few percent higher than those for a
threshold of $\Gamma = 3$, while the number of background
triggers is reduced from 1 to 0. With the implied reduction 
rate in false alarm rate in mind, we choose the CorrPower 
threshold of $\Gamma = 4$.

\begin{figure}[tbh]
        \centering
        \epsfig{file=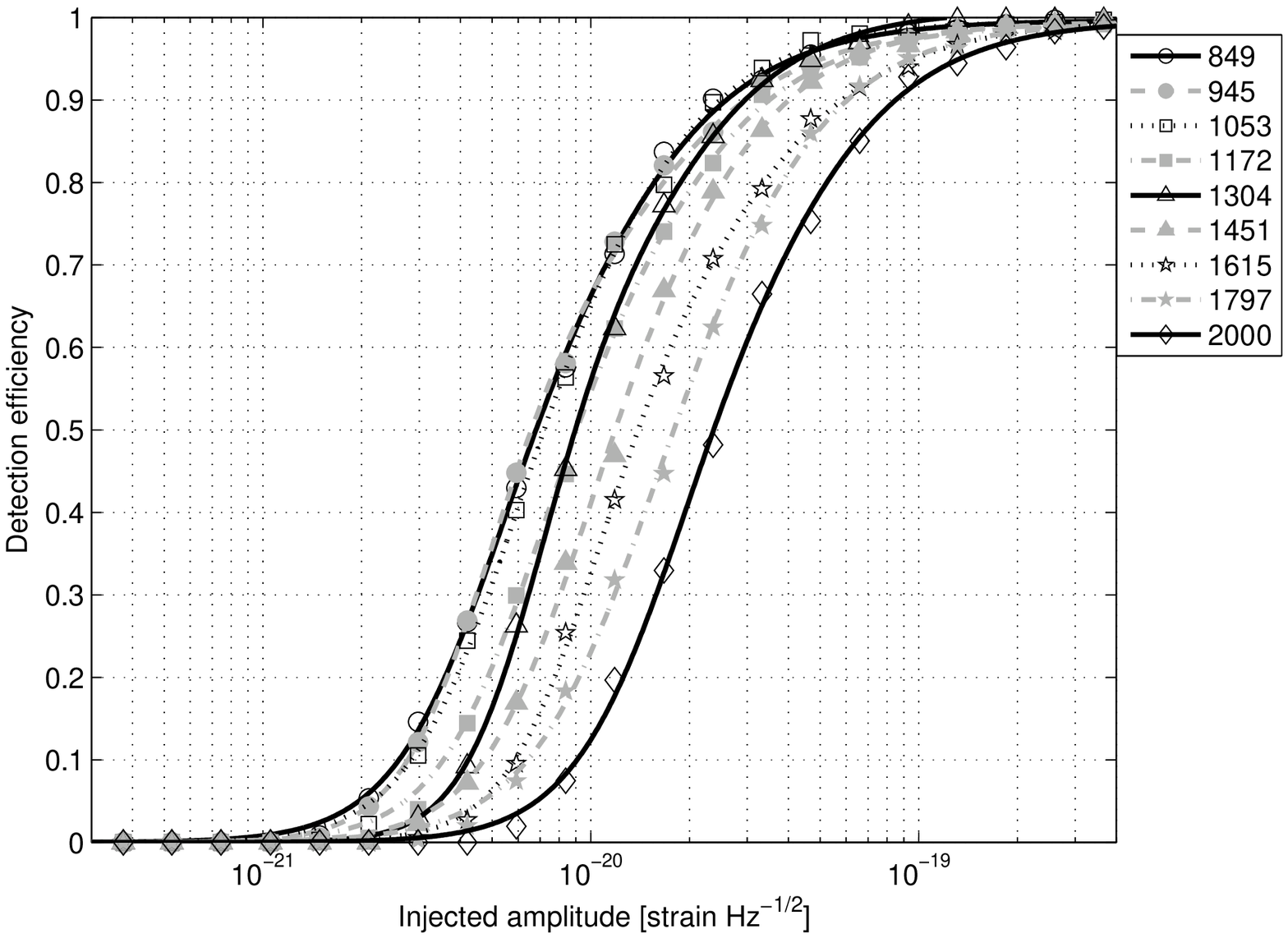, width=\textwidth}
        \caption{Detection efficiency of the WBCP pipeline
	for various sine-Gaussian simulated gravitational-wave
	bursts, as a function of the signal amplitude (defined
	by Eq. 18). The legend indicates the central
	frequency (Hz) of the injected signal. 
        }
        \label{wbcp_eff}
\end{figure}

The fraction of sine-Gaussian signals detected above threshold
(detection efficiency)
as a function of injected $h_\mathrm{rss}$ is shown in Figure
\ref{wbcp_eff}.
Note that the detection efficiencies do not reach 1 for even the
loudest injected signals because of the application of 
auxiliary-channel vetoes. This effect was also observed in~\cite{s4bursts}.
The detector is effectively blind to GW for the duration of
the veto because we are excluding any observations within
this period.
This exclusion means that there is a non-zero false dismissal
probability, even for the loudest GW signals.

\begin{table}
\caption{Table of background triggers and $h^{50\%}_\mathrm{rss}$
as a function of
$\Gamma$. The total number of background triggers observed over
all 100 time-shifts is
shown.}
\begin{tabular}{ccccc}
\hline
 &Number of&\multicolumn{3}{c}{$h^{50\%}_\mathrm{rss}$ [$\times 10^{-21}$ Hz $^{-1/2}$]} \\
$\Gamma$ threshold&background triggers&f = 849\,Hz&1053\,Hz&1615\,Hz\\
\hline
\hline
0&881&6.6&6.9&13.5\\
3&1&6.6&7.1&13.7\\
4&0&6.8&7.2&13.9\\
\hline
\end{tabular}
\label{hrss_gamma}
\end{table}


\subsection{Coherent Waveburst analysis}

\begin{figure}[tbh]
        \centering
        \epsfig{file=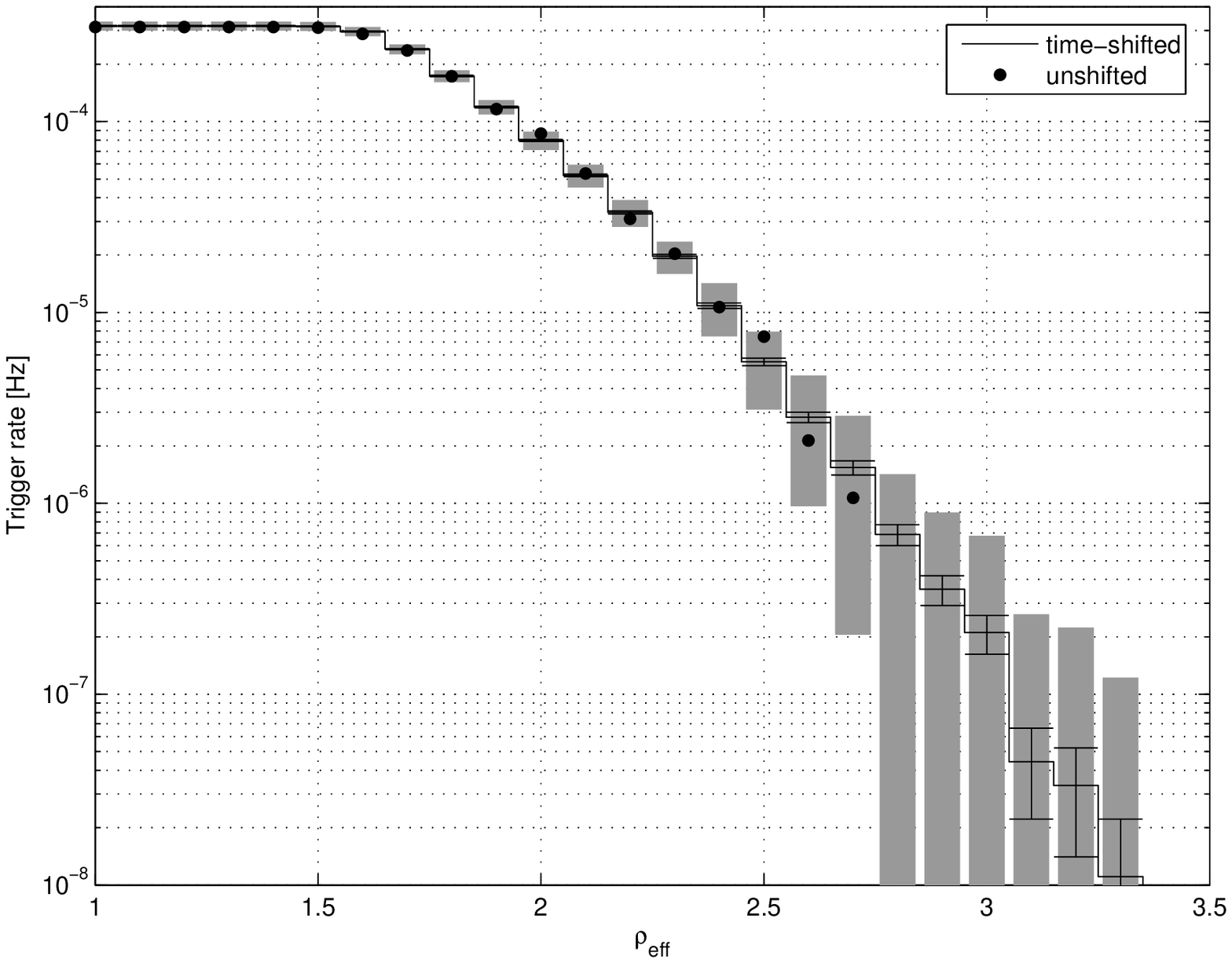, width=\textwidth}
		\caption{Rate of background triggers as a function of effective SNR
		 for the cWB pipeline.
		 The L1 data is shifted in 100 discrete time steps and, 
		 for each threshold value of $\rho_{\rm eff}$, the
		 background rate is calculated by taking an average over
		 all 100 time shifts and plotted as the staircase plot.
		 The $\rho_{\rm eff}$ distribution for unshifted data is 
		 represented by black dots. 
		 As with previous figures, the error bars indicate the
		 range corresponding to $\pm \sqrt{n}/100$, where $n$ is
		 the total number of triggers in each bin.
		 Also, the grey patches indicate the standard deviation in the 
		 number of triggers at each time shift. 
		}
		\label{cwb_2run_rate}
	\end{figure}

For cWB, the tuning strategy is to set thresholds such that no background 
triggers are observed. We first require that $L_{k}$ for all 
three-detector combinations in the network be greater than 36.
We then set the effective SNR threshold high enough to eliminate all
remaining background triggers.
Figure \ref{cwb_2run_rate} shows the quadruple coincidence rate
as a function of the effective SNR, $\rho_{\rm eff}$.
We set a threshold on the effective SNR at 3.4.
This threshold corresponds approximately to the root sum square of the matched
filter SNR of 11 -- 12 detected in the network.

\begin{figure}[tbh]
		\centering
		\epsfig{file=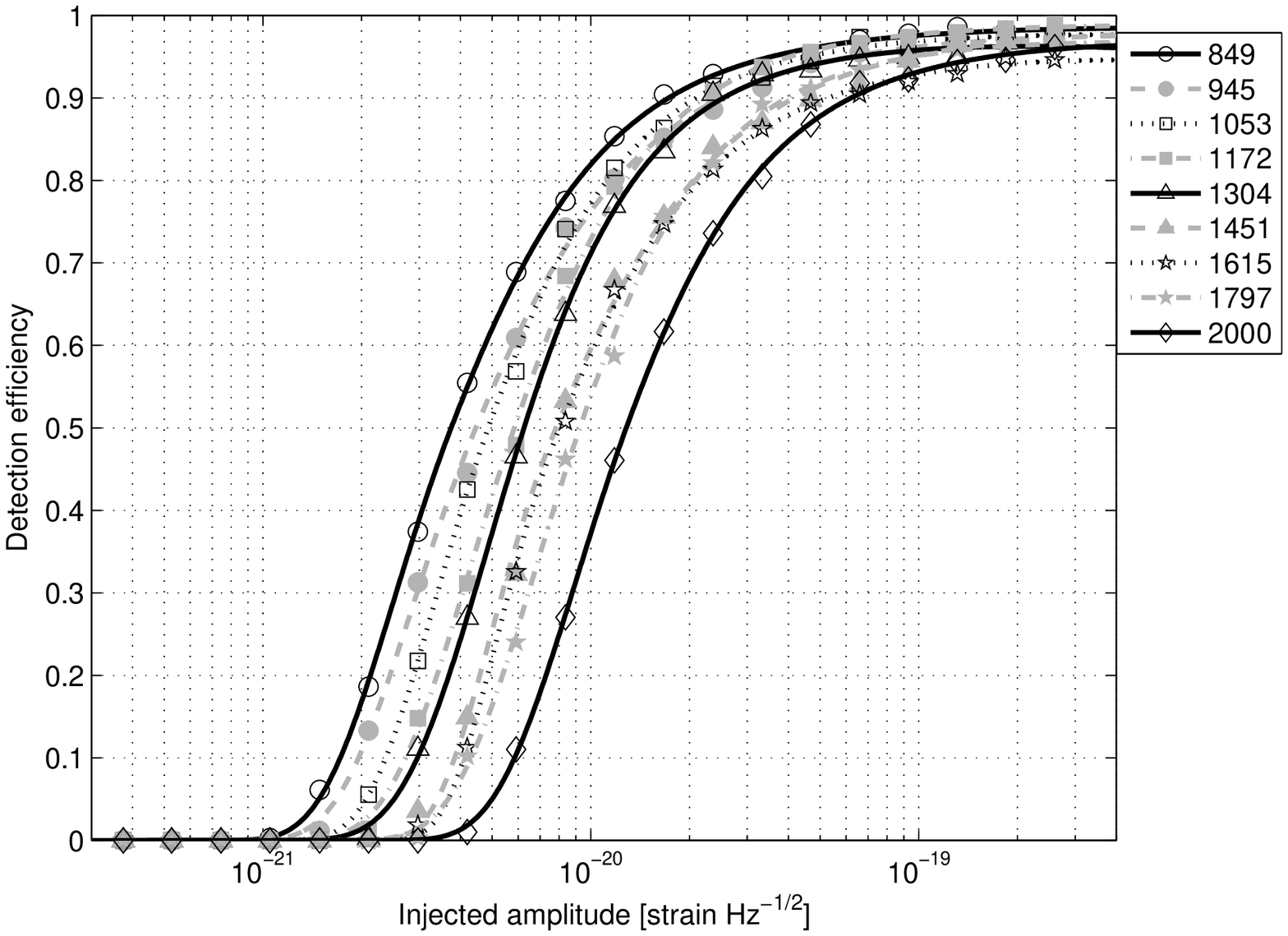, width=\textwidth}
		\caption{Detection efficiency of the coherent Waveburst pipeline
		 for various sine-Gaussian simulated gravitational-wave bursts,
		 as a function of the signal amplitude (defined by Eq. 18).
		 The legend
		 indicates the central frequency (Hz) of the injected
		 signal.
		 }
		\label{cwb_eff}
	\end{figure}

	To determine the detection efficiency, we then inject
	sine-Gaussian burst signals into the data and determine the fraction of 
	injections detected for the selected effective SNR and likelihood thresholds.
	Figure \ref{cwb_eff} plots the detection efficiency as a function
	of the $h_\mathrm{rss}$ of the injected sine-Gaussians. 
	As with the WBCP pipeline, a small fraction of the injection
	signals fall within periods when the data is vetoed. However, in
	addition to this, several injected sine-Gaussians are missed by
	cWB, even at the loudest injection amplitudes, because they have
	sky locations and polarisations where the antenna response at the
	Hanford detector site is very small. This means that the
	injection is missed by both H1 and H2. Of the two remaining
	detectors in the network, the noise in G1 tends to be higher than
	in L1. Therefore, these injected signals are only detected
	strongly by L1 and the trigger does not cross the selected
	thresholds.

\subsection{Zero-lag observations and efficiency comparison}

With the thresholds chosen using the time-shifted analysis
detailed in the previous two subsections, a search for
gravitational waves is performed on LIGO-GEO data between 768
and 2048 Hz with no time shift applied (zero-lag). No
coincidences are observed above the chosen thresholds for
either pipeline.

Figure \ref{rstat-scatter} plots the $\Gamma$ versus $Z_{\rm g}$ scatter 
and $\Gamma$ distribution of the unshifted triggers from the WBCP
pipeline.
From Figure \ref{rstat-scatter}a, it is clear that there are no
unshifted triggers above the pre-determined thresholds of
$\Gamma=4$ and $Z_{\rm g}=5$.
Though the distribution of the unshifted triggers in Figure
\ref{rstat-scatter}b has an outlier at the $\Gamma = 2$ 
histogram bin, one should bear in mind that
these triggers are well below the pre-determined $\Gamma$
threshold of 4. Using the Kolmogorov-Smirnov test,
the statistical significance of the fluctuations in the $\Gamma$
distribution of the unshifted triggers is calculated to be $18\%$,
which means that the null hypothesis is accepted (assuming a
standard significance threshold $5\%$ or greater to accept the
null hypothesis).

The $\rho_{\rm eff}$ distribution
of the unshifted triggers (black dots) for the cWB pipeline is
shown in Figure \ref{cwb_2run_rate}.
The distribution of the unshifted triggers is consistent with the background distribution.
No unshifted triggers were observed above the pre-determined
threshold of $\rho_{\rm eff}=3.4$. In fact, there are no
unshifted triggers with $\rho_{\rm eff} > 2.7$.

\begin{figure}[tbh]
        \centering
        \epsfig{file=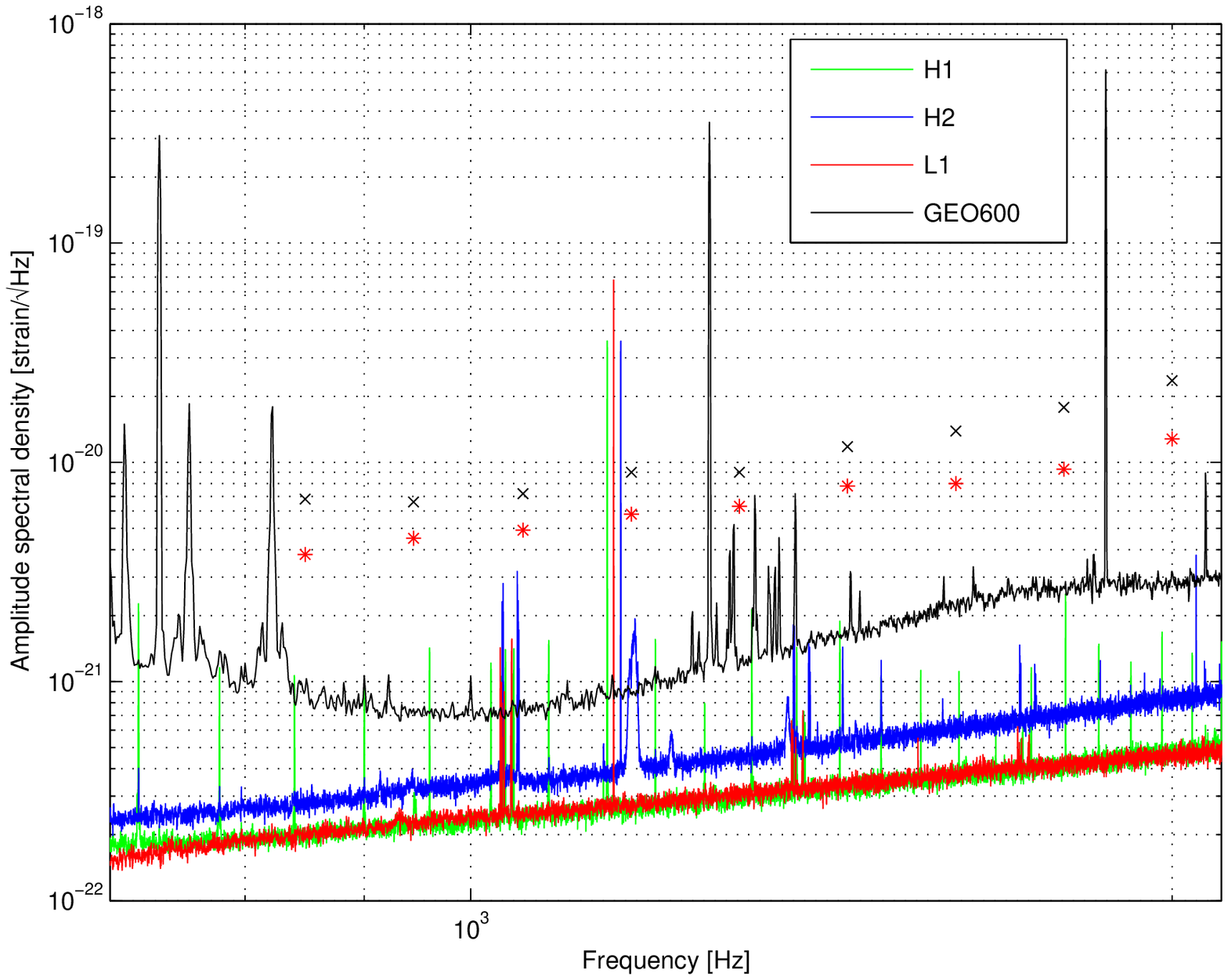, width=\textwidth}
        \caption{The $h_\mathrm{rss}^{50\%}$ values for Waveburst-CorrPower 
        ('x' markers) and coherent Waveburst ('*' markers) pipelines for 
        sine-Gaussians of different central frequencies. 
        Coherent Waveburst is sensitive to gravitational wave signals
        with amplitudes $30-50\%$ lower than those detectable by
        Waveburst-CorrPower.
        } 
        \label{pipeline_eff}
\end{figure}

With no zero-lag coincidences observed in either pipeline, 
we compare the sensitivities of the two pipelines.
We characterise each pipeline's sensitivity by 
the $h_\mathrm{rss}^{50\%}$ values. 
The $h_\mathrm{rss}^{50\%}$ values for the two pipelines used on 
the LIGO-GEO S4 data set are given in Table \ref{eff_comp} and plotted
against the strain spectral densities of the detectors in figure 
\ref{pipeline_eff}. We note that
the $h_\mathrm{rss}^{50\%}$ values obtained for the cWB pipeline are $30-50\%$
lower than those of the WBCP pipeline. 
As desired, the $h_\mathrm{rss}^{50\%}$ values for the cWB 
pipeline are also better than those for the same signals at
these frequencies for a WBCP gravitational-wave burst search
using only LIGO S4 data (4.5 $\times 10^{-21}$ Hz$^{-1/2}$ at 849 Hz and 6.5
$\times 10^{-21}$ Hz$^{-1/2}$ at 1053 Hz)\footnote{This
search was performed in a different frequency range, 64 to 1600
Hz, from that reported here. Additionally, for a fairer comparison, 
the effects of calibration uncertainty have been removed from the values 
quoted here.}~\cite{s4bursts}.

One should also bear in mind that the uncertainty in the
calibration of the detector response to GW has been
conservatively estimated to be $10\%$ for LIGO and
GEO\,600~\cite{calLIGO,calGEO}.
The calibration uncertainty introduces an unknown systematic shifted in
the amplitude scales in Figures \ref{wbcp_eff} and \ref{cwb_eff}.
While the effect of calibration uncertainty is included in the gravitational-wave burst
search with only LIGO S4 data, we have not included calibration uncertainty for the analysis described here.
This is because, while the effect of calibration uncertainty is important for the upper
limits set in \cite{s4bursts}, it is less crucial here since no upper limits have been set.

\begin{table}
\caption{Table of $h^{50\%}_\mathrm{rss}$ as a function of sine-Gaussian central frequencies}
\begin{tabular}{|c|c|c|}
\hline
sine Gaussian&\multicolumn{2}{|c|}{$h^{50\%}_\mathrm{rss}$ [$\times 10^{-21}$ Hz$^{-1/2}$]} \\
\cline{2-3}
central frequency [Hz]&Waveburst-CorrPower&coherent Waveburst \\
\hline
849&6.8&3.8 \\
945&6.6&4.5 \\
1053&7.2&4.9 \\
1172&9.0&5.8 \\
1304&9.0&6.3 \\
1451&11.8&7.8 \\
1615&13.9&8.0 \\
1797&17.8&9.3 \\
2000&23.6&12.8 \\
\hline
\end{tabular}
\label{eff_comp}
\end{table}

\section{Discussion\label{discussion}}

The first joint search for gravitational-wave bursts using the 
LIGO and GEO 600 detectors has been presented.
The search was performed using two pipelines, Waveburst-CorrPower
(WBCP) and coherent Waveburst (cWB), and
targeted signals in the frequency range 768 -- 2048\,Hz. 
No candidate gravitational wave signals have been identified.

The detection efficiencies of the two pipelines to sine-Gaussians
 have been compared. 
The cWB pipeline has $h_{\rm rss}^{50\%}$ values $30-50\%$
lower than those of the WBCP pipeline.
These improved detection efficiencies are also better than those obtained 
for the all-sky burst search using only LIGO S4 data and the WBCP
pipeline~\cite{s4bursts}. One should note, however, that the LIGO-only 
search was
performed at a lower frequency range (64 to 1600 Hz) and optimised for
the characteristics of the noise in that frequency range to maximise
detection efficiency.
Nonetheless, these results show that, for WBCP, the detection efficiency 
is limited by the least sensitive detector when applied to a network of 
detectors with different
antenna patterns and noise levels. This is because WBCP requires that 
excess power be observed in coincidence by all detectors in the network. 
While it is certainly possible to further tune the WBCP pipeline on the
LIGO-GEO S4 data to improve its sensitivity (for example, by reducing
the Waveburst threshold on GEO data or not imposing quadruple coincidence~\cite{LVburst}),
we note that the cWB pipeline naturally includes detectors of
different sensitivities by weighting the data with the antenna
patterns and noise. 
Therefore, with the cWB pipeline, the detection efficiency of the network is
not limited by the least sensitive detector and there is no need for pipeline
tunings that are tailored for particular detector networks.
\\\\

\section*{Acknowledgments}

The authors gratefully acknowledge the support of the United States
National Science Foundation for the construction and operation of the
LIGO Laboratory and the Science and Technology Facilities Council of the
United Kingdom, the Max-Planck-Society, and the State of
Niedersachsen/Germany for support of the construction and operation of
the GEO600 detector. The authors also gratefully acknowledge the support
of the research by these agencies and by the Australian Research Council,
the Council of Scientific and Industrial Research of India, the Istituto
Nazionale di Fisica Nucleare of Italy, the Spanish Ministerio de
Educaci\'on y Ciencia, the Conselleria d'Economia, Hisenda i Innovaci\'o of
the Govern de les Illes Balears, the Royal Society, the Scottish Funding 
Council, the Scottish Universities Physics Alliance, The National
Aeronautics and Space Administration, the Carnegie Trust, the Leverhulme
Trust, the David and Lucile Packard Foundation, the Research Corporation, 
and the Alfred P. Sloan Foundation.
This document has been assigned LIGO Laboratory document number LIGO-P080008-B-Z.
\\
\\

\end{document}